\documentclass[reprint,pra,aps,showpacs,groupedaddress,twocolumn,superscriptaddress]{revtex4-2}

\usepackage[T1]{fontenc}
\usepackage[utf8x]{inputenc}
\usepackage[english]{babel}
\usepackage[usenames,dvipsnames,svgnames]{xcolor}
\usepackage{amsfonts,amsmath,amssymb,stmaryrd}
\usepackage{graphicx}
\usepackage[colorlinks=true,linkcolor=blue,urlcolor=blue,citecolor=blue]{hyperref}

\usepackage[capitalize]{cleveref}  

\newcommand{\qmop}[1]{\boldsymbol{\hat{#1}}}

\newcommand{\expectedvalsimple}[1]{\left\langle #1 \right\rangle}

\newcommand{\ER}{E_\mathrm{R}}

\newcommand{\density}{n}

\newcommand{\avgdensityod}[1][1]{\density_{#1}}
\newcommand{\avglindensity}{\density_{1}}
\newcommand{\kopt}{k_{\mathrm{OL}}}
\newcommand{\lambdaopt}{\lambda_{\mathrm{OL}}}
\newcommand{\VKP}{V_{\mathrm{KP}}}

\newcommand{\scatlenod}{a_{\mathrm{1D}}}
\newcommand{\gfod}{\gf_{ \mathrm{1D} }}

\newcommand{\gf}{g}

\newcommand{\soundvel}[1][\mathrm{s}]{c_{#1}}

\newcommand{\LLgamma}{\gamma}
\newcommand{\LLgasparam}{\avgdensityod \scatlenod}

\newcommand{\densqmop}{\qmop{n}}

\newcommand{\luttparam}{K}
\newcommand{\fermivel}{v_\mathrm{F}}
\newcommand{\fermik}{k_\mathrm{F}}

\newcommand{\KPperiod}{l}
\newcommand{\vzero}{{V_0}}

\newcommand{\ssf}{S}

\newcommand{\latfillfac}{\avgdensityod \KPperiod}
\newcommand{\LLgammac}{\LLgamma_\mathrm{c}}
\newcommand{\luttparamc}{\luttparam_\mathrm{c}}

\begin{document}

\title{Structural Superfluid-Mott Insulator Transition for a Bose Gas in Multi-Rods}
\author{Omar Abel Rodríguez-López}
\email{oarodriguez.mx@gmail.com}
\homepage{https://orcid.org/0000-0002-3635-9248}
\affiliation{Instituto de Física, Universidad Nacional Autónoma de México, Apdo. Postal 20-364, 01000 Ciudad de México, México}

\author{M. A. Solís}
\email{masolis@fisica.unam.mx}
\affiliation{Instituto de Física, Universidad Nacional Autónoma de México, Apdo. Postal 20-364, 01000 Ciudad de México, México}

\author{J. Boronat}
\email{jordi.boronat@upc.edu}
\affiliation{Departament de Física, Universitat Politècnica de Catalunya, Campus Nord B4-B5, E-08034 Barcelona, España}

\date{Modified: \today / Compiled: \today}

\begin{abstract}
  We report on a novel structural Superfluid-Mott Insulator (SF-MI) quantum
  phase transition for an interacting one-dimensional Bose gas within permeable
  multi-rod lattices, where the rod lengths are varied from zero to the lattice
  period length. We use the \emph{ab-initio} diffusion Monte Carlo method to
  calculate the static structure factor, the insulation gap, and the Luttinger
  parameter, which we use to determine if the gas is a superfluid or a Mott
  insulator. For the Bose gas within a \emph{square} Kronig-Penney (KP)
  potential, where barrier and well widths are equal, the SF-MI coexistence
  curve shows the same qualitative and quantitative behavior as that of a
  typical optical lattice with equal periodicity but slightly larger height.
  When we vary the width of the barriers from zero to the length of the
  potential period, keeping the height of the KP barriers, we observe a new way
  to induce the SF-MI phase transition. Our results are of significant interest,
  given the recent  progress on the realization of optical lattices with a
  subwavelength structure that would facilitate their experimental observation.
\end{abstract}

\maketitle

\section{Introduction}

Phase transitions are ubiquitous in condensed matter
physics. In particular, at very low temperatures, quantum phase transitions are
the onset to distinguish new phenomena in quantum many-body systems. Following
the realization of a Bose-Einstein condensate (BEC) in
1995~\cite{bib:anderson-science.269.1995,bib:davis_prl.1995}, a new and
successful research line has been to study BEC within periodic optical lattices.
It has been possible to study
the properties and behavior of atomic gases trapped in three-dimensional (3D) multilayers,
multi-tubes, or a simple cubic array of dots through the superposition of two
opposing lasers in one, two, or three mutually perpendicular
directions~\cite{bib:jaksch_prl.1998,bib:bloch_natphys.2005}, respectively. One
of the most exciting achievements
has been the observation of  the superfluid (SF)
to Mott insulator (MI)
phase transition in a BEC with repulsive interactions, held both in a
3D~\cite{bib:jaksch_prl.1998,bib:greiner_nature.2002} and
one-dimensional (1D)~\cite{bib:haller_nature.2010}
optical lattices.
%

Although there have been great advances in the creation of many types of optical
lattices, efforts are still being made to overcome their limited spatial
resolution, which is of the order of one-half the laser wavelength $\lambdaopt /
  2$, to manipulate atoms. Recently, there has been a notorious interest and
advances in developing tools to surpass the diffraction limit, not only in the
field of cold atoms but also in
nanotechnology~\cite{bib:luo_am.2019,bib:halir_lpr.2015}. Nowadays, the physical
realization of optical lattices
formed by subwavelength (ultranarrow) optical barriers
of width below
$\lambdaopt/50$~\cite{bib:lacki_prl.2016,bib:wang_prl.2018,bib:subhankar_njp.2019,bib:tsui_pra.2020},
is a reality. These so-called subwavelength optical lattices (SWOLs) can be seen
as a very close experimental realization of a sequence of Dirac-$\delta$
functions, forming the well-known Dirac comb
potential~\cite{bib:merzbacher-qm_jws.1969}, and could be useful to test many
mean-field calculations on the weakly-interacting Bose gas in this kind of
potentials~\cite{bib:theodorakis_jpa.1997,bib:weidong_pre.2004,bib:seaman_pra.2005-1,bib:dong_laserphys.2007,bib:rodriguez-lopez_jltp.2020}.
In the near future, we might see the realization of complex subwavelength optical
lattices, including multiscale design,
as a niche to observe new quantum
phenomena~\cite{bib:fedorov_pra.2017,bib:subhankar_njp.2019}.
From the point of view of theoretical simplicity, adaptable periodic structures
ranging from the subwavelength to typical optical lattices could be simulated
and engineered by applying an external Kronig-Penney (KP) potential, for
example, to a quantum gas.

In this paper, we analyze the physical properties of a degenerate
interacting 1D Bose gas in a lattice formed by a succession of permeable rods at
zero temperature. To create this multi-rod lattice, we apply the KP
potential~\cite{bib:kronig_prsa.1931}
\begin{equation}
  \label{eq:kronig-penney-potential}
  \VKP(z) = \vzero \sum_{j=-\infty}^{\infty} \Big(\Theta(z - (j\KPperiod + a)) - \Theta(z - (j + 1)\KPperiod) \Big),
\end{equation}
to the gas, where $\vzero$ is the barrier height and $\Theta (z)$ the Heaviside
step function. The rods, distributed along the $z$ direction, have width $b$,
and are separated by empty regions of length $a$, such that the lattice period
is $\KPperiod \equiv a + b$.
Here, we use the KP potential to study quantum gases
due to its relatively simple shape, robustness, and versatility while preserving
the system essence coming from the periodicity  of its structure.
We use it to model a typical optical lattice $V(z) = \vzero \sin^2(\kopt z)$,
with $\kopt = 2\pi / \lambdaopt$, in the symmetric case $b = a$.
Also,
we model the subwavelength optical barriers in a SWOL through a Kronig-Penney lattice
with $b \ll a$.
%
Furthermore, we take advantage of the KP potential versatility by defining
a new parameter $b/a$, and analyze how its variation affects the ground-state
properties of the Bose gas within a fixed-period multi-rod lattice.
Then, we show a new structural mechanism to induce a reentrant,
commensurate SF-MI-SF quantum phase transition, by varying the ratio $b/a$
while keeping the interaction strength and lattice period fixed. We propose
that this transition, infeasible in experiments with typical optical lattices
under similar conditions, could be observed in experiments with cold atoms
in SWOLs.


\section{Model and theory}

\begin{figure}[t!]
  \centering
  \includegraphics[width=0.975\linewidth]{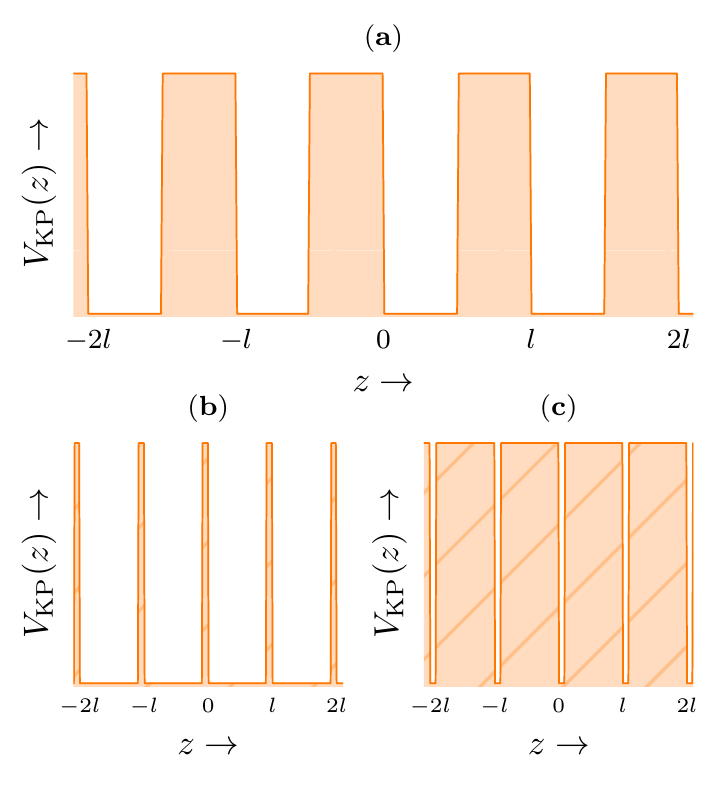}
  \caption{Multi-rod lattices:
    \textbf{(a)} \emph{square} lattice with $b/a = 1$; \textbf{(b)} very thin
    lattice with $b/a = 1/10$; \textbf{(c)} lattice with very broad barriers,
    $b/a = 10$.}%
  \label{fig:figure-1}
\end{figure}

Our analysis relies on the \emph{ab-initio} diffusion Monte Carlo (DMC)
method~\cite{bib:boronat_prb.1994},
adapted for the calculation of pure estimators through the forward-walking
technique. The DMC method solves stochastically the
many-body Schrödinger equation, providing exact results for the ground-state
of the system within some statistical noise.
In \cref{fig:figure-1}, we show three faces of the KP
potential depending on the ratio $b/a$ while keeping $V_0$ and $\KPperiod$
fixed: (a) the symmetric case where $b = a$, which is our reference potential;
(b) when the barriers become very thin, i.e. $b \ll a$; and (c) when $b \gg a$.
The limits $b/a \to 0$ and $b / a \to \infty$ convert the KP potential
in constant potentials $\VKP = 0$ and $\VKP = \vzero$, respectively.
The bosonic particles interact through a contact-like, repulsive
potential of arbitrary magnitude. Consequently, our system corresponds to the
Lieb-Liniger (LL) Bose gas~\cite{bib:lieb_phys-rev.1963} within the multi-rod
lattice~\cref{eq:kronig-penney-potential}. The Hamiltonian of the system with
$N$ bosons is
\begin{equation}
  \label{eq:mb_multrods-ham}
  \hspace{-0.1cm}
  \qmop{H} = -\frac{\hbar^2}{2m} \sum_{i=1}^{N} \left(\frac{\partial^2}{\partial z^2_i} + \VKP(z_i) \right) + \gfod \sum_{i<j}^{N} \delta({z}_{i} - z_j),
\end{equation}
where $m$ is the mass
of the particles, $\gfod \equiv 2\hbar^2 / m \scatlenod$ is the interaction
strength, and $\scatlenod$ is the one-dimensional scattering length. In the
absence of the multi-rod lattice, we recover the LL Bose gas, which is exactly
solvable for any magnitude of the dimensionless interaction parameter $\LLgamma
  \equiv m \gfod
  / \hbar^2 \avgdensityod = 2 / \LLgasparam$, with $\avgdensityod = N/L$ the
linear density.

The presence of a lattice can produce a quantum phase transition in the system
from a superfluid state to a Mott-insulator state or vice-versa, commonly known
as \emph{Mott
  transition}~\cite{bib:greiner_nature.2002,bib:haller_nature.2010,bib:giamarchi.2004}.
In deep optical lattices, that is, when the lattice height is way larger than
the recoil energy $\ER = \hbar^2 \kopt^2 / 2m$, the well-known Bose-Hubbard (BH)
model~\cite{bib:fisher_prb.1989,bib:jaksch_prl.1998,bib:kuhner_prb.1998,bib:greiner_nature.2002,bib:haller_nature.2010,bib:tarruell_crp.2018}
captures the essence of the transition. According to this model, the
competition between the on-site interaction $U$ and the hopping energy
$J$ between adjacent lattice sites drives the transition from the SF to
the MI state. In the Mott insulator state, each lattice site contains the same
number of
particles. On the other hand, the SF-MI transition can be also studied
using the low-energy description of the system given by the Luttinger liquid
theory and the quantum
sine-Gordon (SG)
Hamiltonian~\cite{bib:haldane_prl.1981,bib:giamarchi.2004,bib:cazalilla_rmp.2011,bib:buchler_prl.2003}.
In this approach, two system-dependent quantities, the speed of sound
$\soundvel$ and the Luttinger parameter $\luttparam = \hbar \pi \avgdensityod /
  m \soundvel$, play a central role in the description of the ground-state. In
particular, for any commensurate filling $\avgdensityod \KPperiod
  = {j}/{p}$, with $j$ and $p$ integers, the system can undergo a SF-MI transition
when $\luttparam$ takes the critical value $\luttparam_\mathrm{c} = 2 /
  p^2$~\cite{bib:giamarchi.2004,bib:cazalilla_rmp.2011}, where $p$ is the
commensurability order. For $p=1$, i.e., for an integer number of bosons per
lattice site, $\luttparam_\mathrm{c} = 2$.
The 1D Bose gas remains superfluid
while $\luttparam > \luttparamc$, and variations in the interaction strength and
the lattice height can push $\luttparam$ towards $\luttparamc$. The excitation
energy spectrum is non-gapped and increases linearly with the quasimomentum as
$E(k) = \soundvel \hbar |k|$ in the SF
state~\cite{bib:haldane_jpc.1981,bib:haldane_prl.1981}, whereas it develops an
excitation gap $\Delta$ in the MI phase,  $E(k) = \sqrt{{\left(\soundvel \hbar
        |k| \right)}^2 + \Delta^2}$~\cite{bib:giamarchi.2004}.
Remarkably, in 1D
gases, an arbitrarily weak periodic potential is enough to drive a Mott
transition provided that the interactions are sufficiently
strong~\cite{bib:buchler_prl.2003,bib:haller_nature.2010}.

A fingerprint of the Mott transition can be obtained from the static structure
factor $\ssf(k)$~\cite{bib:yukalov_jps.2007,bib:pitaevskii-stringari_oup.2016},
\begin{equation}
  \label{eq:stat-struct-fac-def}
  \ssf(k) \equiv \frac{1}{N} \left( \expectedvalsimple{\densqmop_{1,k}\densqmop_{1,-k}} - \left|\expectedvalsimple{ \densqmop_{1,k}} \right|^2 \right),
\end{equation}
where the operator $\densqmop_{1,k}$ is the Fourier transform of the density
operator $\densqmop_{1}(z) \equiv \sum_{j=1}^{N} \delta(z - z_j)$. For small
momenta, $S(k)$ is sensitive to the
collective excitations of the system. For high-$k$ values, $S(k)$
approximates to the model-independent value $\lim_{k \to \infty} S(k) = 1$.
Furthermore, the low-momenta behavior of the energy spectrum $E(k)$ is
related to $S(k)$ through the well-known Feynman
relation~\cite{bib:feynman_pr.1954,bib:steinhauer_prl.2002,bib:pitaevskii-stringari_oup.2016},
$E(k) \equiv {\hbar^2 k^2}/{(2 m \ssf(k))}$.
This equation gives an upper bound to the excitation energies in
terms of the static structure factor. Using this expression, we can estimate
both the speed of sound $\soundvel$ and the energy gap $\Delta$ from the
low-momenta behavior
of $\ssf(k)$.
In the SF phase, $\Delta = 0$, and
${(\soundvel / \fermivel)}^{-1} = 2 \fermik \lim_{k \to 0} {S(k)}/{k}$, where
$\fermik = \pi \avglindensity$ is the Fermi momentum and $\fermivel = \hbar
  \fermik   / m$ is the Fermi velocity. Also, $\luttparam = {(\soundvel /
  \fermivel)}^{-1}$.
In contrast, in the MI phase, $S(k)$ grows quadratically with $k$ when $k \to
  0$.

\section{Transition in a Symmetric Lattice}

\begin{figure}[b!]
  \centering
  \includegraphics[width=0.975\linewidth]{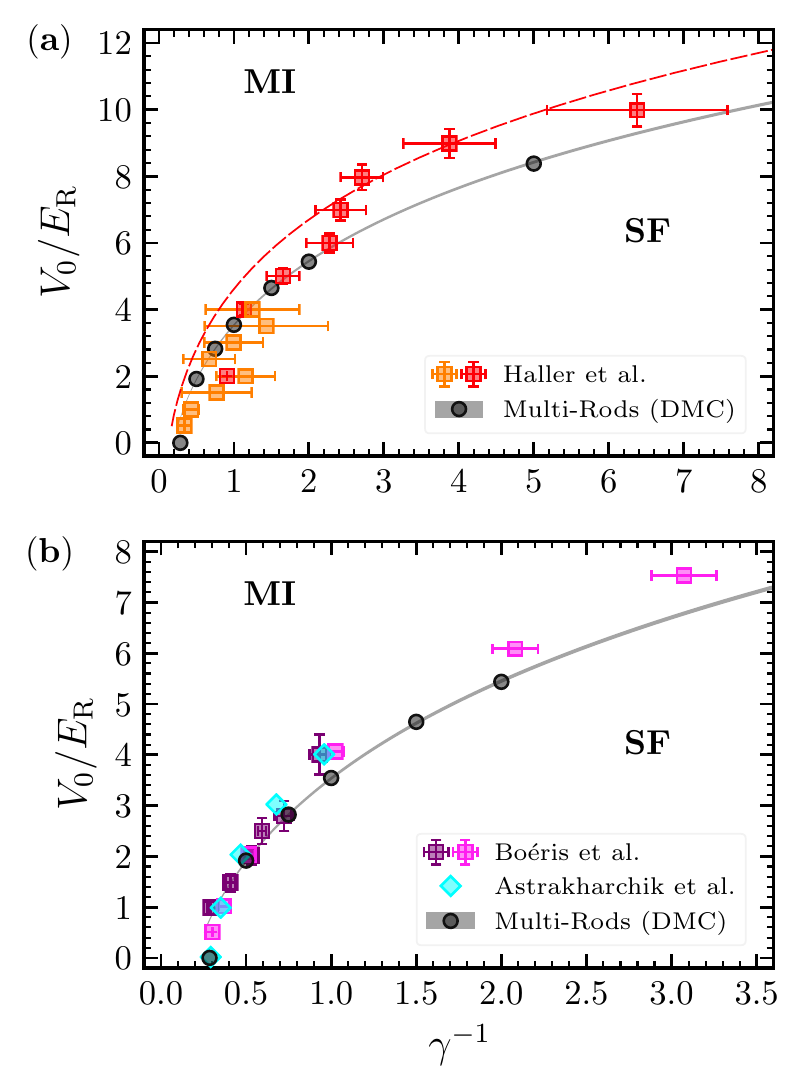}
  \caption{Phase diagram of the 1D Bose gas in a square multi-rods potential at
    zero temperature. The region above the critical points is the Mott-insulator
    (MI) phase, while the region below corresponds to the superfluid (SF) phase.
    (\textbf{a}): We compare our results (black circles) against experimental
    data from Ref.~\cite{bib:haller_nature.2010} (orange and red squares); the
    dashed, red curve is the transition line for an optical lattice according to
    \cref{eq:bh-trans-curve} with ${(U/J)}_\mathrm{c} = 3.85$. The gray line
    is the transition for ${(U/J)}_\mathrm{c} = 2.571(12)$. (\textbf{b}):
    Comparison with data from Ref.~\cite{bib:boeris_pra.2016} (purple and pink
    squares), as well as with simulation data from
    Ref.~\cite{bib:astrakharchik_pra.2016} (blue diamonds). Note that (\textbf{b})
    is an amplification of (\textbf{a}) in the strongly-interacting region.}%
  \label{fig:figure-2}
\end{figure}

We study the SF-MI transition at unit-filling
$\latfillfac = 1$ and $b=a$, i.e.,
our system has, on average, one boson per lattice site. We determine the Mott
transition by calculating $\luttparam$ as a function of the interaction strength
$\LLgamma$ and the lattice height $\vzero$. The boundary between the SF
and MI phases corresponds to the condition $\luttparam(\LLgamma,
  \vzero) = 2$. In \cref{fig:figure-2}a, we show the
DMC zero-temperature phase diagram $\vzero / \ER$ vs. $\LLgamma^{-1}$ for a 1D
Bose gas in a square multi-rod lattice (black circles). In the same figure, we also
show  experimental data~\cite{bib:haller_nature.2010} (orange and red
squares) for a 1D Bose gas in
an optical lattice at commensurability $\avgdensityod \sim 2 / \lambdaopt$,
with $\lambdaopt / 2$  the spatial period of the optical potential.
Note that the recoil energy is $\ER = \hbar^2 \pi^2 / 2 m \KPperiod^2$ since we
require that both optical and multi-rod lattices have the same periodicity,
that is $\KPperiod = \lambdaopt / 2$.

We observe that our results are close to but below the experimental data in the deep lattices zone, where the BH model accurately describes the physics of the
lattice. According to the 1D BH model, for an optical lattice the SF-MI transition occurs at the
critical ratio ${(U/J)}_\mathrm{c} = 3.85$~\cite{bib:rapsch_el.1999}. Since $U =
  (\sqrt{2\pi}/\pi^2) \ER \LLgamma (\avgdensityod \lambdaopt/2) {(\vzero / \ER)}^{1/4}$ and
$J = (4 / \sqrt{\pi}) \ER {(\vzero/\ER)}^{3/4} \exp[-2\sqrt{(\vzero / \ER)}]$,
it is possible to define a relation between the lattice height $\vzero$ and the
interaction strength $\gamma$ at the
transition~\cite{bib:buchler_prl.2003,bib:bloch_rmp.2008},
\begin{equation}
  \label{eq:bh-trans-curve}
  \frac{4\vzero}{\ER} = \ln^2\left[\frac{2\sqrt{2}\pi}{\LLgamma} {\left( \frac{U}{J} \right)}_{\mathrm{c}} \sqrt{\frac{\vzero}{\ER} } \right].
\end{equation}
In \cref{fig:figure-2}a, we show \cref{eq:bh-trans-curve} for the
BH critical value ${(U/J)}_{\mathrm{c}} = 3.85$ (dashed, red line). The
BH transition
line agrees with the experimental data for lattices as high as $\vzero \geq
  7\ER$, but fails for shallow lattices, a behavior discussed
in Ref.~\cite{bib:haller_nature.2010}. Remarkably, most of our DMC simulation
results follow the law \cref{eq:bh-trans-curve}, with a smaller energy
ratio,
${(U/J)}_\mathrm{c} = 2.571(12)$ (dark-gray line), except close to the
transition point $\LLgammac^{-1} = 0.28$ for $\vzero = 0$, where we do not
expect a good fit since the BH model \cref{eq:bh-trans-curve} predict $\gamma^{-1} \to \infty$
when $V_0 \to 0$.
The range of $\vzero$ values for which the BH model fits the multi-rods DMC results is, in
fact, quite large,
starting from
lattice heights as low as $\sim 2 \ER$.
As we can see in \cref{fig:figure-2}a, the square multi-rod lattice is
more insulating than an optical lattice with the same strength.
Looking at the Fourier series expansion of $\VKP(z)$, and considering only up to
second order, we obtain that $\VKP(z) \approx \tilde{V}_0 (1 - \cos(2 \pi z /
  \KPperiod)) / 2 = \tilde{V}_0 \sin^2(\kopt z)$, where $\tilde{V}_0 =
  4\vzero/\pi$. Hence, the multi-rod lattice is roughly equivalent to an optical
lattice with a larger height $\tilde{V}_0$; accordingly, the SF-MI transition in
the former should occur at smaller $\vzero$ than the latter; our
numerical results corroborate this observation.

In \cref{fig:figure-2}b, we focus our analysis on shallow lattices
only, where we do not include the data of Ref.~\cite{bib:haller_nature.2010} to
avoid excessive piling of data.
We compare our results with two additional relevant sources: first,
experimental and numerical data reported in Ref.~\cite{bib:boeris_pra.2016}
(purple and pink squares), and second, DMC data reported in
Ref.~\cite{bib:astrakharchik_pra.2016} (blue diamonds).
Overall, there is a noticeable overlap between our DMC
results and the data from Refs.~\cite{bib:boeris_pra.2016} and~\cite{bib:astrakharchik_pra.2016}, which is stronger as the lattices get shallower.
The
differences between the results for both lattices become smaller as $\LLgammac^{-1} \to
  0.28$ and $\vzero \to
  0$, just as expected since for both multi-rod and optical lattices the system resembles more the
Lieb-Liniger gas.
Experimental data from~\cite{bib:haller_nature.2010}
and~\cite{bib:boeris_pra.2016} in Figs.~\ref{fig:figure-2}a
and~\ref{fig:figure-2}b seem to agree better with the BH model predictions for
multi-rods (gray line) than \cref{eq:bh-trans-curve} for an optical lattice with
${(U/J)}_{\mathrm{c}} = 3.85$ (dashed, red line).
On the other hand, Ref.~\cite{bib:boeris_pra.2016} reports an estimation of
${(U/J)}_\mathrm{c} = 3.36$, for which \cref{eq:bh-trans-curve} agrees better
with experimental data than the BH model for multi-rods. Given the above, it is
clear that the experimental uncertainties in the data reported in
Ref.~\cite{bib:haller_nature.2010} do not allow for a good estimation of
${(U/J)}_\mathrm{c}$ at the transition.

\begin{figure}[b!]
  \centering
  \includegraphics[width=0.975\linewidth]{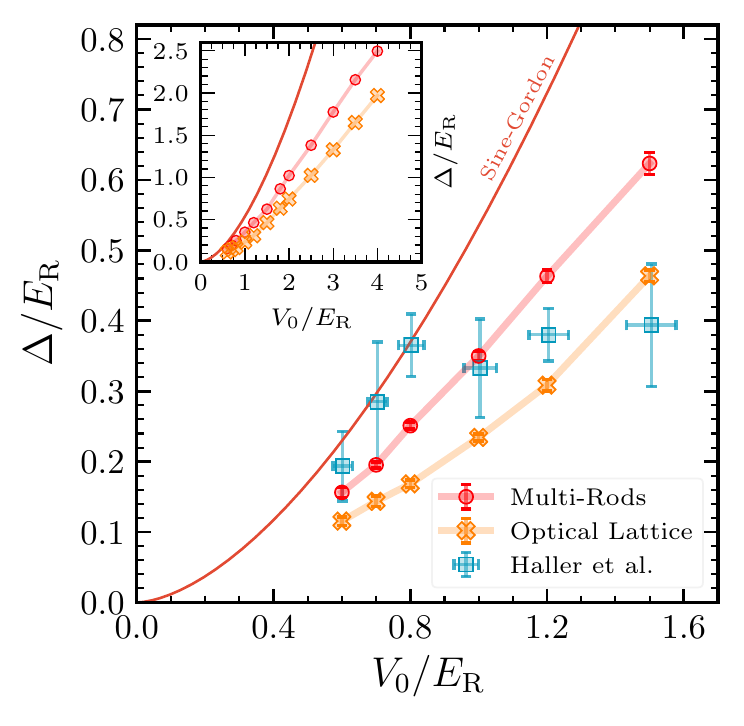}
  \caption{Energy gap for the 1D Bose gas, with $\LLgamma = 11$, in a square
    multi-rod lattice (red circles) and in an optical lattice (orange crosses),
    calculated using DMC.\ Both lattices have the same height $\vzero$. The blue
    squares show experimental gap data from Ref.~\cite{bib:haller_nature.2010}.
    The solid red line indicates the Sine-Gordon
    model~\cite{bib:haller_nature.2010}.
    \textbf{Inset}: gap behavior including lattices with
    $\vzero \geq 1.6\ER$.}%
  \label{fig:figure-3}
\end{figure}

Complementary information on the MI phase can be obtained through the
estimation of the energy gap $\Delta$.  We calculate $\Delta$ as a function
of $\vzero$ for both the multi-rod and optical lattices, with $\LLgamma = 11$.
This particular value of $\LLgamma$ corresponds to the one for available
experimental data~\cite{bib:haller_nature.2010}.
Under these conditions, the system is a strongly interacting gas in the MI phase. We plot our results in \cref{fig:figure-3}, together with the
experimental energy gap reported in Ref.~\cite{bib:haller_nature.2010}. We
observe a good agreement between our simulation results (for both lattices) and
the experimentally measured gap, better than similar results for an optical lattice shown
in Ref.~\cite{bib:astrakharchik_pra.2016}.
However, there is a clear qualitative discrepancy for
$\vzero > 0.8\ER$, since the experimental results show something like a plateau,
while our results grow monotonically. For shallow lattices, the SG model
correctly describes the system's low-energy properties; in particular, it
predicts that $\Delta$ increases with $\vzero$, as our results. As commented
in Ref.~\cite{bib:astrakharchik_pra.2016}, the discussion on how good the
modulation spectroscopy method for measuring $\Delta$ is, remains open. Finally, the gap
as a function of $\vzero$, for an optical lattice is smaller than the gap for a
multi-rod lattice, so the latter is more insulating than the former, confirming the observation made after
analyzing the results shown in \cref{fig:figure-2}.

\section{Structural transition}

\begin{figure}[h!]
  \centering
  \includegraphics[width=0.975\linewidth]{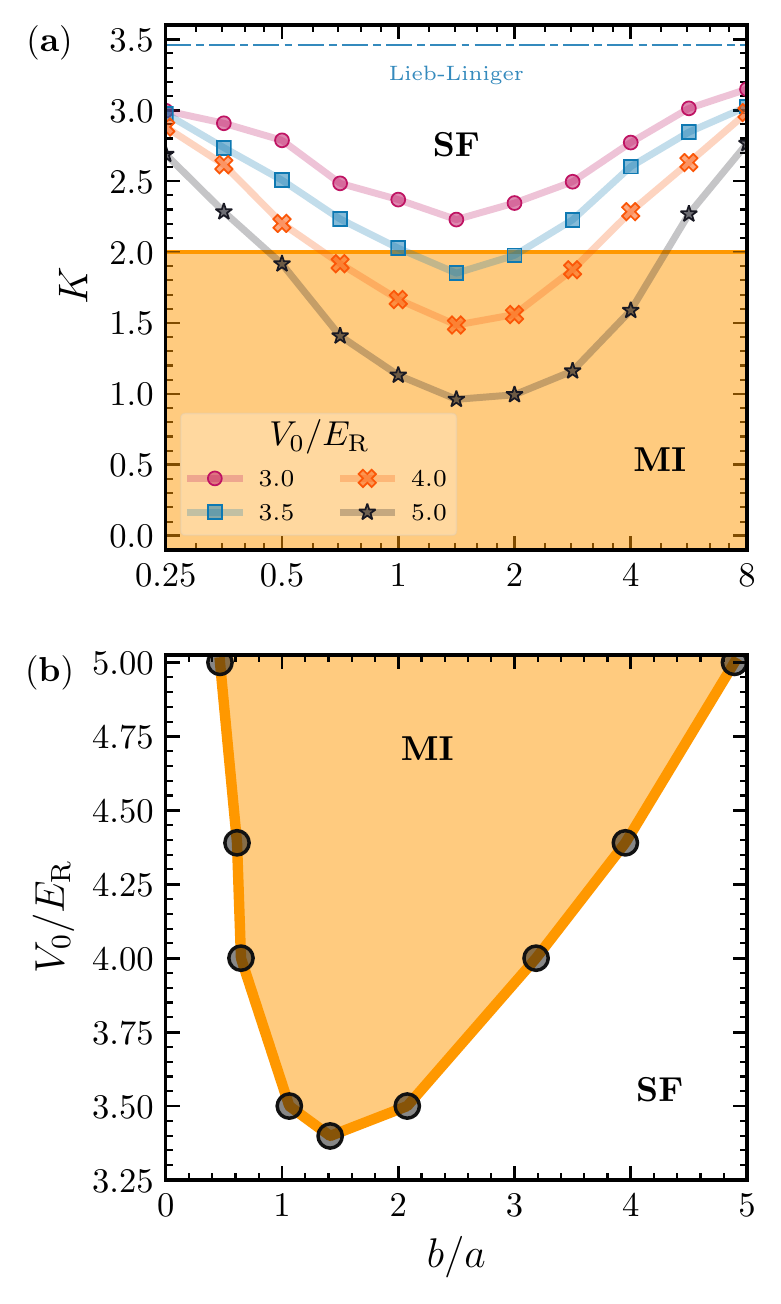}
  \caption{\textbf{(a)}: $\luttparam$ parameter as a function of the lattice
    ratio $b / a$, for $\LLgamma = 1$.
    Independently of $\vzero$, for both $b/a \to 0$ and $b / a \to
      \infty$, $\luttparam$ approaches to the Lieb-Liniger Bose gas $\luttparam =
      3.43$ value. \textbf{(b)}: SF-MI phase transition as a function of $b/a$  and
    $\vzero/\ER$ for $\LLgamma = 1$. The statistical error of the results is
    smaller than the symbol size.}%
  \label{fig:figure-4}
\end{figure}

Motivated by the recent studies on SWOLs, we extend the study of
the interacting Bose gas
within a square multi-rod lattice to a \emph{nonsquare} multi-rod lattice, with special emphasis on describing
the SF-MI quantum phase transition at
commensurability $\avgdensityod \KPperiod = 1$. We performed our analysis for a
fixed interaction strength $\LLgamma = 1$.
We calculate the speed of sound
$\soundvel$ from the low-momenta behavior of the static structure factor, and
determine the $\vzero$ and $b/a$ parameters such that $\luttparam(b/a, \vzero) =
  2$. In \cref{fig:figure-4}a, we show the dependence of the parameter
$\luttparam = {(\soundvel/\fermivel)}^{-1}$ as a function of $b/a$ in four
lattices with heights $\vzero = 3$, $3.5$, $4$, and $5$ times $\ER$.
First, we can see that,
depending on the value of the lattice ratio $b/a$, $\luttparam$ can be greater or
smaller than $\luttparamc = 2$. This result shows that the SF-MI transition can
be triggered by changing the lattices's geometry if its height is large enough.

Variation of
$b/a$ always affects $\luttparam$; however, although $\luttparam$ could
diminish (starting from a SF state), a phase transition may not necessarily
occur for relatively shallow lattices. As one can see, for $b/a \ll 1$
(thin barriers, see \cref{fig:figure-1}b) and $b / a \gg 1$ (thin
wells, see \cref{fig:figure-1}c), the gas becomes superfluid independently of
$\vzero$. On the one hand, as $b / a$ diminishes and the barriers become
thinner, the trapping effect of the lattice greatly reduces; in the limit $b/a
  \to 0$ the system becomes the LL gas. On the other hand, as $b / a$ increases,
the system tends to resemble more to a succession of thin wells; in the limit $b
  / a \to \infty$, it becomes the LL gas subject to a constant potential of height
$\vzero$. Physically, there is no difference between both limits, except by a
shift $\vzero$ in the total energy. Also, in both limits, $K$ approaches to the
corresponding value for the LL model, $K \approx 3.43$.
It is worth noticing that the minimum $\luttparam$ in
\cref{fig:figure-4}a
occurs in the interval $1 < b / a < 2$, which interestingly shows that the
largest trapping effect of the lattice does not correspond to the most symmetric
case, i.e., the square lattice.

We show the zero-temperature phase
transition diagram $\vzero / \ER$ vs. $b/a$ in \cref{fig:figure-4}b.
As commented before, the minimum interaction $\vzero$ to produce the Mott
transition is slightly shifted to non-square potentials, $b/a \simeq 1.4$. When
the asymmetry is $b/a < 1$, the strength $\vzero$ increases quite fast,
favoring the stability of the SF phase. When $b/a > 1.4$, $\vzero$ also
increases but with a slightly smaller slope. The blob in the phase diagram is
therefore not symmetric and interestingly shows the possibility of a double
transition SF-MI-SF for $\vzero > 3.4$ by just changing the relation $b/a$
keeping both $\vzero$ and $\LLgamma$ constants. Note that the \cref{fig:figure-4}b is a cross-section of a MI tubular volume in the $\vzero/\ER$ vs. $\LLgamma^{-1}$ vs. $b/a$ diagram whose asymmetric V-boat-hull shape surface contains the SF-MI phase coexistence line shown in \cref{fig:figure-2}b.

\section{Conclusions}

Using the ab-initio DMC method we calculate the
zero-temperature SF-MI quantum phase transition, in a $\vzero/\ER$ vs.
$\LLgamma^{-1}$ diagram, for a 1D Bose gas with contact interactions within a
square multi-rod lattice.
We show and justify a notable similarity with the phase transition diagram of a
Bose gas in a typical optical lattice by comparing it with several experimental
and numerical data sources, finding that the multi-rod lattice favors the
insulating phase. We also confirm that the BH model accurately predicts the
transition in the regime of weak interactions and deep enough wells.
For the Bose gas in both a multi-rod and optical lattices with the same strength
and period, we calculate the energy gap $\Delta$ as a function of the lattice
height $\vzero$ for $\LLgamma = 11$. As expected, the gap is larger within the
multi-rods lattice than in the optical lattice, validating what has already been
observed in the phase transition diagrams.
In the range of $\vzero$ values where experimental data exist for the gap, our results are
of the same order of magnitude but do not match the
experimental behavior that shows something similar to a saturation with $\vzero$.
%
Finally, we show a new structural mechanism to induce a robust reentrant,
commensurate SF-MI-SF phase transition, triggered by the variation of the
parameter $b/a$ at fixed interaction strength and lattice period. We propose
that such a mechanism could be experimentally implemented using the recently
realized SWOLs, so the SF-MI structural phase transition could be observed.


\begin{acknowledgments}
  We acknowledge partial support from grants PAPIIT-DGAPA-UNAM IN-107616 and
  IN-110319. We also thank the Coordinación de Supercómputo de la Universidad Nacional Autónoma de México for the provided computing resources and technical assistance. This work has been partially supported by the Ministerio de Economia, Industria y Competitividad (MINECO, Spain) under grant No. FIS2017-84114-C2-1-P. We acknowledge financial support from Secretaria d'Universitats i Recerca del Departament d'Empresa i Coneixement de la Generalitat de Catalunya, co-funded by the European Union Regional Development Fund within the ERDF Operational Program of Catalunya (project QuantumCat, ref. 001-P-001644).

\end{acknowledgments}

\bibliographystyle{apsrev4-2}
\bibliography{references}

\end{document}